\documentclass{elsart}
\usepackage{amssymb}
\usepackage{graphics} 

\begin{document}

\begin{frontmatter}

\title{Discrete quasiperiodic sets with predefined local structure}

\author{Nicolae Cotfas}

\address{Faculty of Physics, University of Bucharest,
PO Box 76-54, Post Office 76, Bucharest, Romania}
\ead{ncotfas@yahoo.com}
\begin{abstract}
Model sets play a fundamental role in structure analysis of quasicrystals. 
The diffraction diagram of a quasicrystal
admits as symmetry group a finite group $G$, and there is a $G$-cluster $\mathcal{C}$
(union of orbits of $G$) such that the quasicrystal can be regarded as a 
quasiperiodic packing of interpenetrating copies of $\mathcal{C}$. 
We present an algorithm which leads from any $G$-cluster $\mathcal{C}$ 
directly to a multi-component model set $\mathcal{Q}$ such that the arithmetic
neighbours of any point $x\in \mathcal{Q}$ are distributed on the
sites of the translated copy $x+\mathcal{C}$ of $\mathcal{C}$. 
Our mathematical algorithm may be useful in quasicrystal physics.
\end{abstract}


\end{frontmatter}

\section{Introduction}

Model sets (also called cut-and-project sets) have been introduced by Y. Meyer 
\cite{M} in his study of harmonious sets and later, in the course of the structure 
analysis of quasicrystals, rediscovered in a variety of different schemes 
\cite{KKL,K,KN}.
Extensive investigations \cite{E,H1,H2,K,KN,L1,L2,Me,M,Sch1,Sch2}
 on the properties of these remarkable sets have been 
carried out by Y. Meyer, P. Kramer, M. Duneau, A. Katz, V. Elser, M. Baake, 
R.V. Moody, A. Hof, M. Schlottmann, J.C. Lagarias {\it et. al.}
An extension of the notion of model set called {\it multi-component model set},
very useful in quasicrystal physics, has been introduced by Baake and Moody \cite{B}.
Model sets are generalizations of lattices, and multi-component model sets are
generalizations of lattices with colourings.

Quasicrystals are materials with perfect long-range order, 
but with no three-dimensional translational periodicity.
The structure analysis of quasicrystals on an atomic scale is a highly non-trivial 
task, and we are still far from a satisfactory solution. The electron microscopic images 
suggest the existence of some basic structural units which often overlap (interpenetrate), 
and of some glue atoms. The diffraction spectra contains sharp bright spots,
indicative of long range order, called {\it Bragg reflections}. 
The reflections with intensity above a certain threshold form a discrete set
admitting as symmetry group a finite non-crystallographic group $G$.
In the case of quasicrystals with no translational periodicity this group
is the icosahedral group $Y$ and in the case of quasicrystals 
periodic along one direction (two-dimensional quasicrystals) $G$ is one 
of the dihedral groups $D_8$ (octagonal quasicrystals), $D_{10}$ 
(decagonal quasicrystals) and $D_{12}$ (dodecagonal quasicrystals).

The high resolution microscopic images of a quasicrystal with 
the symmetry group $G$ show that we can regard the quasicrystal as
a quasiperiodic packing of copies of a well-defined $G$-invariant 
finite set $\mathcal{C}$ (basic structural unit), most of them only partially occupied.
From a mathematical point of view, $\mathcal{C}$ is a finite union of orbits of $G$,
and we call it a $G$-{\it cluster}. 
In the literature on quasicrystals the term `cluster' has several meanings \cite{St}.
Depending on the context, it may denote a structure motif (purely geometric pattern),
a structural building block (perhaps with some physical justification), a quasi-unit cell
\cite{SJ} or a complex coordination polyhedron (with some chemical stability).
In our case, $\mathcal{C}$ is a structure motif, perhaps without any physical justification.

The purpose of the present paper is to present a mathematical algorithm which leads
from any $G$-cluster $\mathcal{C}$ directly to a multi-component model set $\mathcal {Q}$ 
representing a quasiperiodic packing of interpenetrating copies of 
$\mathcal{C}$, most of them only partially occupied. 
It shows how to embed the physical space into a superspace
$\mathbb{E}_k$ and how to choose a lattice $\mathbb{L}\subset \mathbb{E}_k$ 
in order to get by projection the desired local structure.
Our algorithm, based on the 
strip projection method and group theory, is a generalization of the model proposed
by Katz and Duneau \cite{K} and independently by Elser \cite{E} for certain icosahedral
quasicrystals. Since the multi-component model
sets have several properties desirable from the physical point of view (they are 
uniformly discrete, relatively dense, have a well-defined density and are
pure point diffractive), our algorithm may be useful in quasicrystal physics.

\section{Model sets and multi-component model sets}

In this section we review some definitions and results concerning the notions of 
model set and multi-component model set.

Let $E$ be a vector subspace of the usual $k$-dimensional Euclidean space  
$\mathbb{E}_k=(\mathbb{R}^k,\langle ,\rangle)$, where
$\langle x,y\rangle =\sum_{i=1}^kx_iy_i$ and $||x||=\sqrt{\langle x,x\rangle }$,
for any $x=(x_i)_{1\leq i\leq k}, \ y=(y_i)_{1\leq i\leq k}$.
The subset $B_r(a)=\{ x\in E\ |\ ||x-a||<r\}$ of $E$,
where $a\in E$, $r\in (0,\infty )$, is the {\em open ball} of center $a$ and radius $r$.\\[2mm]
{\bf Definition 1.} Let $\Lambda $ be a subset of $E$.
\begin{enumerate}
\item The set $\Lambda $ is {\em relatively dense} in $E$ if there is $r\in (0,\infty )$ 
  such that the ball $B_r(x)$ contains at least one point of $\Lambda $, for any $x\in E$.
\item The set $\Lambda $ is {\em uniformly discrete} in $E$ if there is $r\in (0,\infty )$ 
  such that the ball $B_r(x)$ contains at most one point of $\Lambda $, for any $x\in E$.
\item The set $\Lambda $ is a {\em Delone set} in $E$ if $\Lambda $ is both
            relatively dense and uniformly discrete in $E$.
\item The set $\Lambda $ is a {\em lattice} in $E$ if it is both an additive subgroup 
of $E$ and a Delone set in $E$.
\end{enumerate}
{\bf Definition 2.} A {\it cut and project scheme} is a collection of spaces and mappings
\begin{equation}\label{cutproj}
\begin{array}{ccccc}
E_1& \stackrel{\pi _1}\longleftarrow & E_1\oplus E_2
& \stackrel{\pi _2}\longrightarrow & E_2\\
&&\cup && \\
&& L &&
\end{array}
\end{equation} 
formed by two subspaces $E_1$, $E_2$ of $\mathbb{E}_k$,
the corresponding natural projections $\pi _1$, $\pi _2$,
and a lattice $L$ in $E_1\oplus E_2$ such that:
\begin{enumerate}
   \item[(a)] $\pi _1$ restricted to $L$ is one-to-one; 
   \item[(b)] $\pi _2(L)$ is dense in $E_2$.
\end{enumerate}
{\bf Definition 3}. A subset $\Lambda $ of $\mathbb{E}_n$ is a 
{\em regular model set} if there exist
\begin{itemize} 
\item[-] a cut and project scheme (\ref{cutproj}), 
\item[-] an isometry 
$\mathcal{I}:\ \mathbb{E}_n\longrightarrow E_1$ which allows to identify $\mathbb{E}_n$
with $E_1$, 
\item[-] a set $W \not=\emptyset $ satisfying the conditions:
\begin{enumerate}
\item[(i)] $W \subset E_2$ is compact;
\item[(ii)] $W =\overline{{\rm int}(W )}$;
\item[(iii)] The boundary $\partial W$ of $W $ has Lebesgue measure $0$
\end{enumerate}
\end{itemize}
such that 
\begin{equation}
\Lambda =\{ \ \pi _1 x\ |\ x\in L,\ \pi _2x\in W \ \} .
\end{equation}
By using the ${\star }$-mapping 
\begin{equation}
\pi _1(L)\longrightarrow E_2:\ x\mapsto x^{\star }=\pi _2\left( (\pi _1)|_L\right)^{-1}x 
\end{equation}
we can re-write the definition of $\Lambda $ as
\begin{equation}
\Lambda =\{ \ x\ |\ x\in \pi _1(L),\ x^{\star }\in W \ \}.
\end{equation}
Model sets have strong regularity properties.\\[3mm]
{\bf Theorem 1}. \cite{Sch1,Sch2} 
{\it Any regular model set $\Lambda $ is a Delone set and has a well-defined density, 
that is, there exists the limit
\begin{equation}
\lim_{r\rightarrow \infty }\frac{ \#(\Lambda \cap B_r)}{{\rm vol}(B_r)} 
\end{equation}
where $\#(\Lambda \cap B_r)$ is the number of points of $\Lambda $ lying in $B_r=B_r(0)$,
and ${\rm vol}(B_r)$ is the volume of $B_r$.}\\

In structure analysis of quasicrystals, the experimental diffraction image is compared 
with the diffraction image of the mathematical model $\Lambda $, regarded as a set of 
scatterers. In order to compute the diffraction image of the model set $\Lambda $, it
is represented as a Borel measure in the form of a weighted Dirac comb
\begin{equation}
\omega =\sum_{x\in \Lambda }\varphi (x)\delta _x
\end{equation}
where $\varphi : \Lambda \longrightarrow \mathbb{C}$ is a bounded function 
and $\delta _x$ is the Dirac measure located at $x$, that is,
$\delta _x(f)=f(x)$ for continuous functions $f$. 
In this way, atoms of quasicrystal are modeled by their positions and scattering strengths.
In the case (the only considered in the sequel) when there is a function 
$\varrho :E_2\longrightarrow \mathbb{C}$ supported and continuous on $W$ such that
$\varphi (x)=\varrho (x^{\star })$, that is, in the case
$\omega =\sum_{x\in \Lambda }\varrho (x^{\star })\delta _x$
one can prove \cite{H1,H2,B1,B2} the following results:
\begin{itemize}
\item[1.] The measure $\omega $ is translation bounded, that is, there exist constants
          $C_K$ so that
\begin{equation}
\sup_{t\in \mathbb{R}^n}\sum_{x\in \Lambda \cap (t+K)}|\varrho (x)|\leq C_K<\infty 
\end{equation}
for all compact $K\subset \mathbb{R}^n$.
\item[2.] The autocorrelation coefficients 
\begin{equation}
\eta (z)=\lim_{n\rightarrow \infty }\frac{1}{{\rm vol}(B_n)}
\sum_{{\small \begin{array}{c}
x,y\in \Lambda \cap B_n\\
x-y=z
\end{array}}}\varrho (x)\, \overline{\varrho (y)}
\end{equation}
exist for all $z\in \Delta =\Lambda -\Lambda =\{ x-y\ |\ x,y\in \Lambda \}$.
\item[3.] The set $\{ z\in \Delta \ |\ \eta (z)\not=0 \}$ is uniformly discrete.
\item[4.] The autocorrelation measure
\begin{equation}
\gamma _\omega =\sum_{z\in \Delta }\eta (z)\delta _z
\end{equation}
exists.
\end{itemize}

\vspace{3mm}

The diffraction spectrum of $\Lambda $ (the idealized mathematical interpretation
of the diffraction pattern of a physical experiment)
is related \cite{H1,H2} to the Fourier
transform $\hat{\gamma }_\omega $ of the autocorrelation measure $\gamma _\omega $
which can be decomposed as 
\begin{equation}
\hat{\gamma }_\omega =(\hat{\gamma }_\omega )_{pp}+
                      (\hat{\gamma }_\omega )_{sc}+(\hat{\gamma }_\omega )_{ac}
\end{equation}
by the Lebesgue decomposition theorem. Here $\hat{\gamma }_\omega (B)$ is the total
intensity scattered into the volume $B$, $(\hat{\gamma }_\omega )_{pp}$ is a pure
point measure, which corresponds to the Bragg part of the diffraction spectrum,
$(\hat{\gamma }_\omega )_{ac}$ is absolutely continuous and 
$(\hat{\gamma }_\omega )_{sc}$
singular continuous with respect to Lebesgue measure. 
We say that $\Lambda $ is {\it pure point diffractive} if 
$\hat{\gamma }_\omega =(\hat{\gamma }_\omega )_{pp}$, that is , if 
$(\hat{\gamma }_\omega )_{sc}=(\hat{\gamma }_\omega )_{ac}=0$.
We have the following result.\\[3mm]
{\bf Theorem 2} \cite{H1,Sch2} {\it Regular model sets are pure point 
diffractive.}\\[3mm]
In the case of certain model sets used in quasicrystal physics as a mathematical 
model, the agreement between theoretic and experimental diffraction image 
is rather good \cite{E,K}.\\ 

The notion of model set admits the following generalization \cite{B}.\\[3mm]
{\bf Definition 4}. A subset $\Lambda $ of $\mathbb{E}_n$ is an 
{\em $m$-component model set} (also called a {\em multi-component model set}) if there exist
\begin{itemize} 
\item[-] a cut and project scheme (\ref{cutproj}), 
\item[-] an isometry 
$\mathcal{I}:\ \mathbb{E}_n\longrightarrow E_1$ which allows to identify $\mathbb{E}_n$
with $E_1$, 
\item[-] a lattice $M$ in $E_1\oplus E_2$ containing $L$ as a sublattice,
\item[-] $m$ cosets $L_j=\theta _j+L$ of $L$ in $M$,\\ where $j\in \{ 1,2,...,m\}$, 
\item[-] $m$ sets $W _j$ satisfying (i)-(iii), where $j\in \{ 1,2,...,m\}$,
\end{itemize}
such that 
\begin{equation}
\Lambda =\bigcup_{j=1}^m\{ \pi _1 x\ |\ x\in L_j,\ \pi _2x\in W _j\} .
\end{equation}
{\bf Theorem 3.} \cite{B} 
{\it Any multi-component model set is a Delone set, has a well-defined 
density and is pure point diffractive.}

The multi-component model sets have the property of finite local complexity, that is,
there are only finitely many translational classes of clusters of $\Lambda $ with any 
given size. The orbit of $\Lambda $ under translation gives rise, via completion in
the standard Radin-Wolff type topology, to a compact space $X_\Lambda $, and one 
obtains a dynamical system $(X_\Lambda ,\mathbb{R}^n)$. The connection existing 
between the spectrum of this dynamical system and the diffraction measure allows one
to use of some powerful spectral theorems in the study of multi-component model sets
\cite{LMS1,LMS2}.

\section{Model sets with predefined local structure}

Let $\{ g:\mathbb{E}_n\longrightarrow \mathbb{E}_n\ |\ g\in G\}$ be a faithful
orthogonal $\mathbb{R}$-irreducible representation of a finite group $G$, and let
\begin{equation}
{\mathcal C}=\bigcup_{x\in S}Gx\cup \bigcup_{x\in S}G(-x)
              =\{ e_1,...,e_k,-e_1,...,-e_k\}
\end{equation}
be the $G$-cluster symmetric with respect to the origin generated by a finite set 
$S\subset \mathbb{E}_n$. For each $g\in G,$ there exist
the numbers $s_1^g,\ s_2^g,...,s_k^g\in \{ -1;\ 1\}$ and a permutation of 
the set $\{ 1,2,...,k\}$ denoted also by $g$ such that
\begin{equation}\label{sp}
ge_j=s_{g(j)}^ge_{g(j)}\qquad {\rm for\ all\ }
j\in \{ 1,2,...,k\}. 
\end{equation}
Let $e_i=(e_{ij})_{1\leq j\leq n}$, and let
$\varepsilon _i=(\delta_{ij})_{1\leq j\leq k}$, where $\delta _{ij}=1$ for 
$i=j$, and $\delta _{ij}=0$ for $i\not=j$.\\[3mm]
{\bf Lemma 1.} \cite{C1,C2} {\it  
The formula $g\varepsilon _j=s_{g(j)}^g\varepsilon _{g(j)}$
defines the orthogonal representation
\begin{equation}
g(x_i)_{1\leq i\leq k}=\left(s_i^gx_{g^{-1}(i)}\right)_{1\leq i\leq k}
\end{equation}
of $G$ in $\mathbb{E}_k.$ The subspace
\begin{equation}
{\bf E}=\left\{ \ (<u,e_i>)_{1\leq i\leq k}\ |\ \ u\in \mathbb{E}_n\ \right\}
\end{equation}
of $\mathbb{E}_k$ is $G$-invariant and the vectors 
\[ w_j=\kappa ^{-1} (e_{ij})_{1\leq i\leq k}\qquad j\in \{ 1,2,...,n\} \]
where $ \kappa =\sqrt{(e_{11})^2+(e_{21})^2+...+(e_{k1})^2}$, 
form an orthonormal basis of ${\bf E}$.}\\[3mm]
{\bf Lemma 2.} \cite{C1,C2} {\it 
a) \ The subduced representation of $G$ in ${\bf E}$
is equivalent with the representation of $G$ in $\mathbb{E}_n,$ and the isomorphism
of representations 
\begin{equation}
{\mathcal I}:\mathbb{E}_n\longrightarrow {\bf E}\qquad 
\mathcal{I}u=\left(\kappa ^{-1} \langle u,e_i\rangle \right)_{1\leq i\leq k}
\end{equation}
with the property 
${\mathcal I}(\alpha _1,\alpha _2,...,\alpha _n)
=\alpha _1 w_1+...+\alpha _n w_n$ allows us to identify 
the `physical' space $\mathbb{E}_n$ with the subspace ${\bf E}$ of $\mathbb{E}_k$.\\[2mm]
b) \ The matrix of the orthogonal projector 
$\pi \!:\!\mathbb{E}_k\!\longrightarrow \!\mathbb{E}_k$ corresponding to ${\bf E}$ in the basis
$\{ \varepsilon _1,\varepsilon _2,...,\varepsilon _k\}$  is
\begin{equation}
\pi =\left(\kappa ^{-2}\langle e_i,e_j\rangle \right)_{1\leq i,j\leq k}
\end{equation}
c) \ The lattice \ 
$\mathbb{L}=\kappa \mathbb{Z}^k\subset \mathbb{E}_k$ \
is \ $G$-invariant, \ $\pi (\kappa \varepsilon _i)={\mathcal I}e_i$,
that is, $\pi (\kappa \varepsilon _i)=e_i$ if we take into consideration 
the identification ${\mathcal I}:\mathbb{E}_n\longrightarrow {\bf E},$ and
\begin{equation}
\pi (\mathbb{L})=\mathbb{Z}e_1+\mathbb{Z}e_2+...+\mathbb{Z}e_k.
\end{equation}
 }

Let $\mathcal{V}$ be a $G$-invariant subspace of $\mathbb{E}_k$.
Since the representation of $G$ in $\mathbb{E}_k$ is orthogonal and
$\langle gx,y\rangle =\langle gx,g(g^{-1})y\rangle =\langle x,g^{-1}y\rangle $
the orthogonal complement 
\begin{equation}
\mathcal{V}^\perp =\{ x\in \mathbb{E}_k\ |\ \langle x,y\rangle =0\ 
{\rm for\ all}\ y\in \mathcal{V}\}
\end{equation}
of $\mathcal{V}$ is also a $G$-invariant subspace.
The orthogonal projectors $\Pi ,\, \Pi ^\perp :\mathbb{E}_k\longrightarrow \mathbb{E}_k$
corresponding to $\mathcal{V}$ and $\mathcal{V}^\perp $ satisfy the relations
\[
\begin{array}{lll}
\Pi \circ \Pi =\Pi  & \Pi \circ \Pi ^\perp =0 & \Pi \circ g=g\circ \Pi \\
\Pi ^\perp \circ \Pi ^\perp =\Pi ^\perp \quad & \Pi ^\perp \circ \Pi =0 \quad &
\Pi ^\perp \circ g=g\circ \Pi ^\perp 
\end{array}
\]
for any $g\in G$.\\[3mm]
{\bf Theorem 4.} {\it 
\begin{itemize}
\item[a)] If $\Pi (\mathbb{L})$ is dense in $\mathcal{V}$ then 
         $\mathbb{L}\cap \mathcal{V}=\{ 0\}$.
\item[b)] If $\Pi (\mathbb{L})$ is discrete in $\mathcal{V}$ then 
   $\mathbb{L}\cap \mathcal{V}$ contains a basis of $\mathcal{V}$.
\item[c)] If $\Pi (\mathbb{L})$ is discrete in $\mathcal{V}$ then 
   $\Pi ^\perp (\mathbb{L})$ is a lattice in $\mathcal{V}^\perp $. 
\end{itemize} }

{\bf Proof.} a) Let us assume that there is $z\in \mathbb{L}\cap \mathcal{V}$, 
$z\not=0$. For each $y\in \mathbb{L}$ the
solutions $x=(x_1,x_2,...,x_k)$ of the equation
\[ z_1(x_1-y_1)+z_2(x_2-y_2)+...+z_k(x_k-y_k)=0\]
form the hyperplane $H_y$ orthogonal to $z$ passing through $y$. 
The hyperplane $H_y$ intersect the one-dimensional subspace 
$\mathbb{R}z=\{ \alpha z\ |\ \alpha \in \mathbb{R}\}$
at a point corresponding to $\alpha =\langle y,z\rangle /||z||^2$.
Since $\langle y,z\rangle \in \kappa ^2\mathbb{Z}$, the minimal distance between
two distinct hyperplanes of the family of parallel hyperplanes 
$\{ H_y\ |\ y\in \mathbb{L}\}$ containing $\Pi (\mathbb{L})$ is 
$\kappa ^2/||z||$. The set  $\Pi (\mathbb{L})$ which is contained in 
union $H=\bigcup_{y\in \mathbb{L}}H_y$ can not be dense in $\mathcal{V}$. 
Each point of $\mathbb{R}z-H$ belongs to $\mathcal{V}$ but can not be 
the limit of a sequence of points from $\Pi (\mathbb{L})$.

b) In view of a well-known result \cite{D} concerning lattices in subspaces of
$\mathbb{E}_k$, there exist $\lambda _1$, $\lambda _2$, ... , $\lambda _s$ in
$\mathbb{L}$ such that $\{ \Pi \lambda _1, \Pi \lambda _2,...,
\Pi \lambda _s\}$ is a basis in $\mathcal {V}$ and 
$\Pi (\mathbb{L})=\mathbb{Z}\Pi \lambda _1+
                      \mathbb{Z}\Pi \lambda _2+...
                      \mathbb{Z}\Pi \lambda _s .$
We extend $\{ \lambda _1, \lambda _2, ... , \lambda _s\}$ up to a basis 
$\{ \lambda _1, \lambda _2, ... , \lambda _k\}$ of $\mathbb{E}_k$ by adding new
vectors $\lambda _{s+1}$, ... , $\lambda _k$ from $\mathbb{L}$. 
For each $i\in \{ s+1,s+2,...,k\}$ there are $\alpha _{i1}$, ... ,
$\alpha _{is}\in \mathbb{Z}$ such that 
\[ \Pi \lambda _i=\alpha _{i1}\Pi \lambda _1+\alpha _{i2}\Pi \lambda _2+
                 ...+\alpha _{is}\Pi \lambda _s \]
that is,
\[ \Pi (\lambda _i-\alpha _{i1}\lambda _1-\alpha _{i2}\lambda _2-
                 ...-\alpha _{is}\lambda _s)=0. \]
The linearly independent vectors
\[ v_i=\lambda _i-\alpha _{i1}\lambda _1-
                 ...-\alpha _{is}\lambda _s\qquad i\in \{ s+1,s+2,...,k\} \]
belonging to $\mathbb{L}$ form a basis in $\mathcal{V}^\perp $.
Since the coordinates $v_{i1}$, $v_{i2}$,...,$v_{ik}$ of each vector
$v_i$ belong to $\kappa \mathbb{Z}$ the space $\mathcal{V}$ which coincides to the
space of all the solutions $x=(x_1,x_2,...,x_k)$ of the system of linear equations
\[ v_{i1}x_1+v_{i2}x_2+...+v_{ik}x_k=0\qquad i\in \{ s+1,s+2,...k\} \]
contains $s$ linearly independent vectors $v_1$, $v_2$, ... , $v_s$ from $\mathbb{L}$.
They form a basis of $\mathcal{V}$.

c) The vectors $v_1$, $v_2$, ... , $v_k$ from $\mathbb{L}$ form a basis of $\mathbb{E}_k$.
Since 
\[ \Pi ^\perp v_i=\left\{ \begin{array}{lll}
0 & {\rm for} & i\in \{ 1,2,...,s\}\\
v_i & {\rm for} & i\in \{ s+1,s+2,...,k\}
\end{array} \right. \]
and the change of basis matrix from $\{ \kappa \varepsilon _1, \kappa \varepsilon _2,
...,\kappa \varepsilon _k\}$ to $\{ v_1,v_2,...,v_k\}$ has rational entries it 
follows that the entries of $\Pi $ in the basis 
$\{ \kappa \varepsilon _1,...,\kappa \varepsilon _k\}$ 
are rational. 
If $q$ is the least common multiple of the denominators of the entries of $\Pi ^\perp $
then $\Pi ^\perp (\mathbb{L})$ is contained in the discrete set 
$(\kappa /q)\mathbb{Z}^k$.\qquad  \rule{2mm}{2mm}\\

In order to obtain a description of the structure of $\mathbb{Z}$-module 
$\Pi (\mathbb{L})$ we use the following result.\\[3mm]
{\bf Theorem 5.} \cite{D,S} {\it Let $\phi :\, \mathbb{R}^k\longrightarrow \mathbb{R}^l$
be a surjective linear mapping, where $l<k$. Then there are subspaces
$V_1,\, V_2$ of $\mathbb{R}^l$ such that:
\begin{enumerate}
\item[a)] $\mathbb{R}^l=V_1\oplus V_2$ 
\item[b)] $\phi (\mathbb{Z}^k)=\phi (\mathbb{Z}^k)\cap V_1+
                                \phi (\mathbb{Z}^k)\cap V_2$,
\item[c)] $\phi (\mathbb{Z}^k)\cap V_2$ is a lattice in $V_2$, 
\item[d)] $\phi (\mathbb{Z}^k)\cap V_1$ is a dense subgroup of $V_1$.
\end{enumerate}
The subspace $V_1$ in this decomposition is uniquely determined.}\\[3mm]
{\bf Theorem 6.} {\it If $\mathcal{V}$ is a $G$-invariant subspace
of $\mathbb{E}_k$ and if $\Pi $ is the corresponding orthogonal
projector then there exist two subspaces $\mathcal{V}_1$, $\mathcal{V}_2$ such that:
\begin{enumerate}
\item[a)] $\mathcal{V}=\mathcal{V}_1\oplus \mathcal{V}_2$,
\item[b)] $\Pi (\mathbb{L})=\Pi (\mathbb{L})\cap \mathcal{V}_1+
                                \Pi (\mathbb{L})\cap \mathcal{V}_2$,
\item[c)] $\Pi (\mathbb{L})\cap \mathcal{V}_2$ is a lattice in 
           $\mathcal{V}_2$, 
\item[d)] $\Pi (\mathbb{L})\cap \mathcal{V}_1$ is a $\mathbb{Z}$-module dense
           in $\mathcal{V}_1$.
\end{enumerate}
The subspace $\mathcal{V}_1$ is uniquely determined and $G$-invariant}.\\ 

{\bf Proof.} The existence of the decomposition follows directly from the 
previous theorem. It remains only to prove the $G$-invariance of $\mathcal{V}_1$.
For each $x\in \mathcal{V}_1$ there is a sequence 
$(\xi _j)_{j\geq 0}$ in $\Pi (\mathbb{L})$
such that $x=\lim_{j\rightarrow \infty }\xi _j$ and $\xi _j\not=x$ for all $j$. 
The transformation 
$g:\mathbb{E}_k\longrightarrow \mathbb{E}_k$ corresponding to each $g\in G$ is an
isometry and $g(\mathbb{L})=\mathbb{L}$. Therefore 
\[ \begin{array}{l}
gx=\lim_{j\rightarrow \infty }g\xi _j\qquad g\xi _j\not=gx\\[2mm]
g(\Pi (\mathbb{L}))=\Pi (g(\mathbb{L}))=\Pi (\mathbb{L})
\end{array} \]
whence $gx\in \mathcal{V}_1$.\qquad  \rule{2mm}{2mm}

\begin{figure}
\setlength{\unitlength}{0.7mm}
\begin{picture}(130,60)(-10,0)
\put(80,20){$\mathcal{C}$}
\put(147,10){$\mathcal{Q}$}
\put(62,19){{\scriptsize ${\bf E}$}}
\put(-1,19){{\scriptsize ${\bf E}'$}}
\put(8,42){{\scriptsize ${\bf E}''$}}
\put(24.5,43){{\scriptsize ${\bf E}^\perp $}}
\put(18.2,37){{\scriptsize ${\bf W}$}}
\put(7,32.5){{\scriptsize ${\bf W_i}$}}
\put(7,14.8){{\scriptsize ${\bf W_j}$}}
\put(57,26){{\scriptsize ${\mathcal E}$}}
\put(32,31){ {\scriptsize ${\mathcal E}_i=z_i+{\mathcal E}$}}
\put(32,13){{\scriptsize ${\mathcal E}_j=z_j+{\mathcal E}$}}
\put(4,0){\line(0,1){38}}
\put(4,0){\line(3,1){20}}
\put(4,38){\line(3,1){20}}
\put(24,6.8){\line(0,1){38}}
\put(4,19){\line(1,0){40}}
\put(24,25.5){\line(1,0){40}}
\put(44,19){\line(3,1){20}}
\put(4,28){\line(1,0){40}}
\put(4,28){\line(3,1){20}}
\put(24,34.5){\line(1,0){40}}
\put(44,28){\line(3,1){20}}
\put(4,10){\line(1,0){40}}
\put(4,10){\line(3,1){20}}
\put(24,16.5){\line(1,0){40}}
\put(44,10){\line(3,1){20}}
\put(9,5){\line(4,1){10}}
\put(9,5){\line(-1,4){4.3}}
\put(4.8,22.4){\line(1,4){3.5}}
\put(18.8,7.3){\line(1,4){4}}
\put(22.5,23.4){\line(-1,3){5}}
\put(8.7,36.4){\line(4,1){8.5}}
\linethickness{0.4mm}
\put(14,22.3){\line(1,0){50}}
\put(0,17.5){\line(3,1){24.1}}
\put(0,17.4){\line(3,1){24.1}}
\put(0,17.6){\line(3,1){24.1}}
\put(5.7,19.2){\line(3,1){16.4}}
\put(5.7,19.7){\line(3,1){16.4}}
\put(6.5,29){\line(3,1){12.8}}
\put(6.5,29.1){\line(3,1){12.8}}
\put(6.5,29.2){\line(3,1){12.8}}
\put(6.5,28.9){\line(3,1){12.8}}
\put(6.5,28.8){\line(3,1){12.8}}
\put(7.7,11.4){\line(3,1){13}}
\put(7.7,11.5){\line(3,1){13}}
\put(7.7,11.6){\line(3,1){13}}
\put(7.7,11.2){\line(3,1){13}}
\put(7.7,11.3){\line(3,1){13}}
\linethickness{0.4mm}
\put(14,3.5){\line(0,1){42}}
\setlength{\unitlength}{2mm}
\put(   44.30000,  10.72426){\circle*{0.2}} 
\put(   43.30000,  10.72426){\circle*{0.2}} 
\put(   43.59289,  10.01716){\circle*{0.2}} 
\put(   45.00711,  11.43137){\circle*{0.2}} 
\put(   44.30000,   9.72426){\circle*{0.2}} 
\put(   44.30000,  11.72426){\circle*{0.2}} 
\put(   45.00711,  10.01716){\circle*{0.2}} 
\put(   42.59289,  10.01716){\circle*{0.2}} 
\put(   43.30000,  11.72426){\circle*{0.2}} 
\put(   42.59289,  11.43137){\circle*{0.2}} 
\put(   43.59289,   9.01716){\circle*{0.2}} 
\put(   46.00711,  11.43137){\circle*{0.2}} 
\put(   45.00711,  12.43137){\circle*{0.2}} 
\put(   45.71421,  10.72426){\circle*{0.2}} 
\put(   45.00711,   9.01716){\circle*{0.2}} 
\put(   43.59289,  12.43137){\circle*{0.2}} 
\put(   46.00711,  10.01716){\circle*{0.2}} 
\put(   41.59289,  10.01716){\circle*{0.2}} 
\put(   42.59289,   9.01716){\circle*{0.2}} 
\put(   41.88579,  10.72426){\circle*{0.2}} 
\put(   42.59289,  12.43137){\circle*{0.2}} 
\put(   44.30000,   8.31005){\circle*{0.2}} 
\put(   46.00711,  12.43137){\circle*{0.2}} 
\put(   46.71421,  10.72426){\circle*{0.2}} 
\put(   45.71421,  13.13848){\circle*{0.2}} 
\put(   45.00711,  13.43137){\circle*{0.2}} 
\put(   44.30000,  13.13848){\circle*{0.2}} 
\put(   46.00711,   9.01716){\circle*{0.2}} 
\put(   45.71421,   8.31005){\circle*{0.2}} 
\put(   41.59289,   9.01716){\circle*{0.2}} 
\put(   40.88579,  10.72426){\circle*{0.2}} 
\put(   41.88579,   8.31005){\circle*{0.2}} 
\put(   43.30000,   8.31005){\circle*{0.2}} 
\put(   41.88579,  11.72426){\circle*{0.2}} 
\put(   41.59289,  12.43137){\circle*{0.2}} 
\put(   43.30000,  13.13848){\circle*{0.2}} 
\put(   42.59289,  13.43137){\circle*{0.2}} 
\put(   41.88579,  13.13848){\circle*{0.2}} 
\put(   44.30000,   7.31005){\circle*{0.2}} 
\put(   45.00711,   7.60294){\circle*{0.2}} 
\put(   46.71421,  13.13848){\circle*{0.2}} 
\put(   46.71421,  11.72426){\circle*{0.2}} 
\put(   47.71421,  10.72426){\circle*{0.2}} 
\put(   47.42132,  11.43137){\circle*{0.2}} 
\put(   46.71421,   9.72426){\circle*{0.2}} 
\put(   47.42132,  10.01716){\circle*{0.2}} 
\put(   45.71421,  14.13848){\circle*{0.2}} 
\put(   44.30000,  14.13848){\circle*{0.2}} 
\put(   46.71421,   8.31005){\circle*{0.2}} 
\put(   40.88579,   8.31005){\circle*{0.2}} 
\put(   40.88579,   9.72426){\circle*{0.2}} 
\put(   40.17868,  10.01716){\circle*{0.2}} 
\put(   40.88579,  11.72426){\circle*{0.2}} 
\put(   40.17868,  11.43137){\circle*{0.2}} 
\put(   41.88579,   7.31005){\circle*{0.2}} 
\put(   42.59289,   7.60294){\circle*{0.2}} 
\put(   43.30000,   7.31005){\circle*{0.2}} 
\put(   40.88579,  13.13848){\circle*{0.2}} 
\put(   43.30000,  14.13848){\circle*{0.2}} 
\put(   41.88579,  14.13848){\circle*{0.2}} 
\put(   43.59289,   6.60294){\circle*{0.2}} 
\put(   45.00711,   6.60294){\circle*{0.2}} 
\put(   46.00711,   7.60294){\circle*{0.2}} 
\put(   47.71421,  13.13848){\circle*{0.2}} 
\put(   46.71421,  14.13848){\circle*{0.2}} 
\put(   47.42132,  12.43137){\circle*{0.2}} 
\put(   48.42132,  11.43137){\circle*{0.2}} 
\put(   48.42132,  10.01716){\circle*{0.2}} 
\put(   47.42132,   9.01716){\circle*{0.2}} 
\put(   45.00711,  14.84558){\circle*{0.2}} 
\put(   43.59289,  14.84558){\circle*{0.2}} 
\put(   47.71421,   8.31005){\circle*{0.2}} 
\put(   46.71421,   7.31005){\circle*{0.2}} 
\put(   40.17868,   7.60294){\circle*{0.2}} 
\put(   40.88579,   7.31005){\circle*{0.2}} 
\put(   40.17868,   9.01716){\circle*{0.2}} 
\put(   39.17868,  10.01716){\circle*{0.2}} 
\put(   39.47157,  10.72426){\circle*{0.2}} 
\put(   40.17868,  12.43137){\circle*{0.2}} 
\put(   39.17868,  11.43137){\circle*{0.2}} 
\put(   42.59289,   6.60294){\circle*{0.2}} 
\put(   40.88579,  14.13848){\circle*{0.2}} 
\put(   42.59289,  14.84558){\circle*{0.2}} 
\put(   44.30000,   5.89584){\circle*{0.2}} 
\put(   46.00711,   6.60294){\circle*{0.2}} 
\put(   45.00711,   5.60294){\circle*{0.2}} 
\put(   47.71421,  14.13848){\circle*{0.2}} 
\put(   48.42132,  12.43137){\circle*{0.2}} 
\put(   47.42132,  14.84558){\circle*{0.2}} 
\put(   46.00711,  14.84558){\circle*{0.2}} 
\put(   49.12843,  10.72426){\circle*{0.2}} 
\put(   48.42132,   9.01716){\circle*{0.2}} 
\put(   45.00711,  15.84558){\circle*{0.2}} 
\put(   44.30000,  15.55269){\circle*{0.2}} 
\put(   47.71421,   7.31005){\circle*{0.2}} 
\put(   48.42132,   7.60294){\circle*{0.2}} 
\put(   47.42132,   6.60294){\circle*{0.2}} 
\put(   39.17868,   7.60294){\circle*{0.2}} 
\put(   40.17868,   6.60294){\circle*{0.2}} 
\put(   39.47157,   8.31005){\circle*{0.2}} 
\put(   41.59289,   6.60294){\circle*{0.2}} 
\put(   39.17868,   9.01716){\circle*{0.2}} 
\put(   38.47157,  10.72426){\circle*{0.2}} 
\put(   39.17868,  12.43137){\circle*{0.2}} 
\put(   40.17868,  13.43137){\circle*{0.2}} 
\put(   39.47157,  13.13848){\circle*{0.2}} 
\put(   41.88579,   5.89584){\circle*{0.2}} 
\put(   42.59289,   5.60294){\circle*{0.2}} 
\put(   43.30000,   5.89584){\circle*{0.2}} 
\put(   41.59289,  14.84558){\circle*{0.2}} 
\put(   40.17868,  14.84558){\circle*{0.2}} 
\put(   43.30000,  15.55269){\circle*{0.2}} 
\put(   42.59289,  15.84558){\circle*{0.2}} 
\put(   41.88579,  15.55269){\circle*{0.2}} 
\put(   44.30000,   4.89584){\circle*{0.2}} 
\put(   46.00711,   5.60294){\circle*{0.2}} 
\put(   46.71421,   5.89584){\circle*{0.2}} 
\put(   45.71421,   4.89584){\circle*{0.2}} 
\put(   48.42132,  14.84558){\circle*{0.2}} 
\put(   48.42132,  13.43137){\circle*{0.2}} 
\put(   49.42132,  12.43137){\circle*{0.2}} 
\put(   49.12843,  13.13848){\circle*{0.2}} 
\put(   49.12843,  11.72426){\circle*{0.2}} 
\put(   47.42132,  15.84558){\circle*{0.2}} 
\put(   46.71421,  15.55269){\circle*{0.2}} 
\put(   46.00711,  15.84558){\circle*{0.2}} 
\put(   50.12843,  10.72426){\circle*{0.2}} 
\put(   49.12843,   9.72426){\circle*{0.2}} 
\put(   49.42132,   9.01716){\circle*{0.2}} 
\put(   49.12843,   8.31005){\circle*{0.2}} 
\put(   45.71421,  16.55269){\circle*{0.2}} 
\put(   44.30000,  16.55269){\circle*{0.2}} 
\put(   48.42132,   6.60294){\circle*{0.2}} 
\put(   39.17868,   6.60294){\circle*{0.2}} 
\put(   38.47157,   8.31005){\circle*{0.2}} 
\put(   40.17868,   5.60294){\circle*{0.2}} 
\put(   40.88579,   5.89584){\circle*{0.2}} 
\put(   38.47157,   9.72426){\circle*{0.2}} 
\put(   37.47157,  10.72426){\circle*{0.2}} 
\put(   37.76447,  10.01716){\circle*{0.2}} 
\put(   38.47157,  11.72426){\circle*{0.2}} 
\put(   37.76447,  11.43137){\circle*{0.2}} 
\put(   38.47157,  13.13848){\circle*{0.2}} 
\put(   39.47157,  14.13848){\circle*{0.2}} 
\put(   41.88579,   4.89584){\circle*{0.2}} 
\put(   43.30000,   4.89584){\circle*{0.2}} 
\put(   40.88579,  15.55269){\circle*{0.2}} 
\put(   39.17868,  14.84558){\circle*{0.2}} 
\put(   40.17868,  15.84558){\circle*{0.2}} 
\put(   43.30000,  16.55269){\circle*{0.2}} 
\put(   41.88579,  16.55269){\circle*{0.2}} 
\put(   43.59289,   4.18873){\circle*{0.2}} 
\put(   45.00711,   4.18873){\circle*{0.2}} 
\put(   46.71421,   4.89584){\circle*{0.2}} 
\put(   47.71421,   5.89584){\circle*{0.2}} 
\put(   49.42132,  14.84558){\circle*{0.2}} 
\put(   49.12843,  15.55269){\circle*{0.2}} 
\put(   48.42132,  15.84558){\circle*{0.2}} 
\put(   49.12843,  14.13848){\circle*{0.2}} 
\put(   50.12843,  13.13848){\circle*{0.2}} 
\put(   50.12843,  11.72426){\circle*{0.2}} 
\put(   46.71421,  16.55269){\circle*{0.2}} 
\put(   50.83553,  11.43137){\circle*{0.2}} 
\put(   50.12843,   9.72426){\circle*{0.2}} 
\put(   50.83553,  10.01716){\circle*{0.2}} 
\put(   50.12843,   8.31005){\circle*{0.2}} 
\put(   49.12843,   7.31005){\circle*{0.2}} 
\put(   45.00711,  17.25980){\circle*{0.2}} 
\put(   44.30000,  17.55269){\circle*{0.2}} 
\put(   43.59289,  17.25980){\circle*{0.2}} 
\put(   49.42132,   6.60294){\circle*{0.2}} 
\put(   48.42132,   5.60294){\circle*{0.2}} 
\put(   49.12843,   5.89584){\circle*{0.2}} 
\put(   38.47157,   5.89584){\circle*{0.2}} 
\put(   39.17868,   5.60294){\circle*{0.2}} 
\put(   38.47157,   7.31005){\circle*{0.2}} 
\put(   37.47157,   8.31005){\circle*{0.2}} 
\put(   37.76447,   9.01716){\circle*{0.2}} 
\put(   39.47157,   4.89584){\circle*{0.2}} 
\put(   40.88579,   4.89584){\circle*{0.2}} 
\put(   36.76447,  10.01716){\circle*{0.2}} 
\put(   36.76447,  11.43137){\circle*{0.2}} 
\put(   37.76447,  12.43137){\circle*{0.2}} 
\put(   37.47157,  13.13848){\circle*{0.2}} 
\put(   38.47157,  14.13848){\circle*{0.2}} 
\put(   42.59289,   4.18873){\circle*{0.2}} 
\put(   40.88579,  16.55269){\circle*{0.2}} 
\put(   39.17868,  15.84558){\circle*{0.2}} 
\put(   38.47157,  15.55269){\circle*{0.2}} 
\put(   39.47157,  16.55269){\circle*{0.2}} 
\put(   42.59289,  17.25980){\circle*{0.2}} 
\put(   43.59289,   3.18873){\circle*{0.2}} 
\put(   44.30000,   3.48162){\circle*{0.2}} 
\put(   46.00711,   4.18873){\circle*{0.2}} 
\put(   45.00711,   3.18873){\circle*{0.2}} 
\put(   47.71421,   4.89584){\circle*{0.2}} 
\put(   47.42132,   4.18873){\circle*{0.2}} 
\put(   50.12843,  15.55269){\circle*{0.2}} 
\put(   50.12843,  14.13848){\circle*{0.2}} 
\put(   49.12843,  16.55269){\circle*{0.2}} 
\put(   47.71421,  16.55269){\circle*{0.2}} 
\put(   50.83553,  12.43137){\circle*{0.2}} 
\put(   47.42132,  17.25980){\circle*{0.2}} 
\put(   46.71421,  17.55269){\circle*{0.2}} 
\put(   46.00711,  17.25980){\circle*{0.2}} 
\put(   51.83553,  11.43137){\circle*{0.2}} 
\put(   51.54264,  10.72426){\circle*{0.2}} 
\put(   50.83553,   9.01716){\circle*{0.2}} 
\put(   51.83553,  10.01716){\circle*{0.2}} 
\put(   50.12843,   7.31005){\circle*{0.2}} 
\put(   50.83553,   7.60294){\circle*{0.2}} 
\put(   45.00711,  18.25980){\circle*{0.2}} 
\put(   43.59289,  18.25980){\circle*{0.2}} 
\put(   50.12843,   5.89584){\circle*{0.2}} 
\put(   49.12843,   4.89584){\circle*{0.2}} 
\put(   37.47157,   5.89584){\circle*{0.2}} 
\put(   38.47157,   4.89584){\circle*{0.2}} 
\put(   37.76447,   6.60294){\circle*{0.2}} 
\put(   37.47157,   7.31005){\circle*{0.2}} 
\put(   36.76447,   7.60294){\circle*{0.2}} 
\put(   36.76447,   9.01716){\circle*{0.2}} 
\put(   40.17868,   4.18873){\circle*{0.2}} 
\put(   41.59289,   4.18873){\circle*{0.2}} 
\put(   36.05736,  10.72426){\circle*{0.2}} 
\put(   36.76447,  12.43137){\circle*{0.2}} 
\put(   37.47157,  14.13848){\circle*{0.2}} 
\put(   37.76447,  14.84558){\circle*{0.2}} 
\put(   42.59289,   3.18873){\circle*{0.2}} 
\put(   41.59289,  17.25980){\circle*{0.2}} 
\put(   40.88579,  17.55269){\circle*{0.2}} 
\put(   40.17868,  17.25980){\circle*{0.2}} 
\put(   38.47157,  16.55269){\circle*{0.2}} 
\put(   42.59289,  18.25980){\circle*{0.2}} 
\put(   44.30000,   2.48162){\circle*{0.2}} 
\put(   46.00711,   3.18873){\circle*{0.2}} 
\put(   46.71421,   3.48162){\circle*{0.2}} 
\put(   45.71421,   2.48162){\circle*{0.2}} 
\put(   48.42132,   4.18873){\circle*{0.2}} 
\put(   47.42132,   3.18873){\circle*{0.2}} 
\put(   50.12843,  16.55269){\circle*{0.2}} 
\put(   50.83553,  14.84558){\circle*{0.2}} 
\put(   50.83553,  13.43137){\circle*{0.2}} 
\put(   49.12843,  17.55269){\circle*{0.2}} 
\put(   48.42132,  17.25980){\circle*{0.2}} 
\put(   51.83553,  12.43137){\circle*{0.2}} 
\put(   51.54264,  13.13848){\circle*{0.2}} 
\put(   47.42132,  18.25980){\circle*{0.2}} 
\put(   46.00711,  18.25980){\circle*{0.2}} 
\put(   52.54264,  10.72426){\circle*{0.2}} 
\put(   51.83553,   9.01716){\circle*{0.2}} 
\put(   51.54264,   8.31005){\circle*{0.2}} 
\put(   50.83553,   6.60294){\circle*{0.2}} 
\put(   45.71421,  18.96690){\circle*{0.2}} 
\put(   44.30000,  18.96690){\circle*{0.2}} 
\put(   50.12843,   4.89584){\circle*{0.2}} 
\put(   37.47157,   4.89584){\circle*{0.2}} 
\put(   36.76447,   6.60294){\circle*{0.2}} 
\put(   37.76447,   4.18873){\circle*{0.2}} 
\put(   39.17868,   4.18873){\circle*{0.2}} 
\put(   36.05736,   8.31005){\circle*{0.2}} 
\put(   35.76447,   9.01716){\circle*{0.2}} 
\put(   36.05736,   9.72426){\circle*{0.2}} 
\put(   40.17868,   3.18873){\circle*{0.2}} 
\put(   40.88579,   3.48162){\circle*{0.2}} 
\put(   41.59289,   3.18873){\circle*{0.2}} 
\put(   35.05736,  10.72426){\circle*{0.2}} 
\put(   36.05736,  11.72426){\circle*{0.2}} 
\put(   35.76447,  12.43137){\circle*{0.2}} 
\put(   36.76447,  13.43137){\circle*{0.2}} 
\put(   36.05736,  13.13848){\circle*{0.2}} 
\put(   36.76447,  14.84558){\circle*{0.2}} 
\put(   37.76447,  15.84558){\circle*{0.2}} 
\put(   41.88579,   2.48162){\circle*{0.2}} 
\put(   43.30000,   2.48162){\circle*{0.2}} 
\put(   41.59289,  18.25980){\circle*{0.2}} 
\put(   40.17868,  18.25980){\circle*{0.2}} 
\put(   39.17868,  17.25980){\circle*{0.2}} 
\put(   37.47157,  16.55269){\circle*{0.2}} 
\put(   38.47157,  17.55269){\circle*{0.2}} 
\put(   37.76447,  17.25980){\circle*{0.2}} 
\put(   43.30000,  18.96690){\circle*{0.2}} 
\put(   41.88579,  18.96690){\circle*{0.2}} 
\put(   44.30000,   1.48162){\circle*{0.2}} 
\put(   45.00711,   1.77452){\circle*{0.2}} 
\put(   46.71421,   2.48162){\circle*{0.2}} 
\put(   49.42132,   4.18873){\circle*{0.2}} 
\put(   48.42132,   3.18873){\circle*{0.2}} 
\put(   49.12843,   3.48162){\circle*{0.2}} 
\put(   50.83553,  17.25980){\circle*{0.2}} 
\put(   50.12843,  17.55269){\circle*{0.2}} 
\put(   50.83553,  15.84558){\circle*{0.2}} 
\put(   51.83553,  14.84558){\circle*{0.2}} 
\put(   51.54264,  14.13848){\circle*{0.2}} 
\put(   48.42132,  18.25980){\circle*{0.2}} 
\put(   52.54264,  13.13848){\circle*{0.2}} 
\put(   52.54264,  11.72426){\circle*{0.2}} 
\put(   46.71421,  18.96690){\circle*{0.2}} 
\put(   53.54264,  10.72426){\circle*{0.2}} 
\put(   53.24975,  11.43137){\circle*{0.2}} 
\put(   52.54264,   9.72426){\circle*{0.2}} 
\put(   53.24975,  10.01716){\circle*{0.2}} 
\put(   52.54264,   8.31005){\circle*{0.2}} 
\put(   51.54264,   7.31005){\circle*{0.2}} 
\put(   51.83553,   6.60294){\circle*{0.2}} 
\put(   50.83553,   5.60294){\circle*{0.2}} 
\put(   51.54264,   5.89584){\circle*{0.2}} 
\put(   45.00711,  19.67401){\circle*{0.2}} 
\put(   44.30000,  19.96690){\circle*{0.2}} 
\put(   50.83553,   4.18873){\circle*{0.2}} 
\put(   36.76447,   4.18873){\circle*{0.2}} 
\put(   36.76447,   5.60294){\circle*{0.2}} 
\put(   35.76447,   6.60294){\circle*{0.2}} 
\put(   36.05736,   5.89584){\circle*{0.2}} 
\put(   36.05736,   7.31005){\circle*{0.2}} 
\put(   37.76447,   3.18873){\circle*{0.2}} 
\put(   38.47157,   3.48162){\circle*{0.2}} 
\put(   39.17868,   3.18873){\circle*{0.2}} 
\put(   35.05736,   8.31005){\circle*{0.2}} 
\put(   35.05736,   9.72426){\circle*{0.2}} 
\put(   39.47157,   2.48162){\circle*{0.2}} 
\put(   40.88579,   2.48162){\circle*{0.2}} 
\put(   34.35025,  10.01716){\circle*{0.2}} 
\put(   35.05736,  11.72426){\circle*{0.2}} 
\put(   34.35025,  11.43137){\circle*{0.2}} 
\put(   35.05736,  13.13848){\circle*{0.2}} 
\put(   36.05736,  14.13848){\circle*{0.2}} 
\put(   35.76447,  14.84558){\circle*{0.2}} 
\put(   36.76447,  15.84558){\circle*{0.2}} 
\put(   36.05736,  15.55269){\circle*{0.2}} 
\put(   41.88579,   1.48162){\circle*{0.2}} 
\put(   42.59289,   1.77452){\circle*{0.2}} 
\put(   43.30000,   1.48162){\circle*{0.2}} 
\put(   40.88579,  18.96690){\circle*{0.2}} 
\put(   39.17868,  18.25980){\circle*{0.2}} 
\put(   39.47157,  18.96690){\circle*{0.2}} 
\put(   36.76447,  17.25980){\circle*{0.2}} 
\put(   37.76447,  18.25980){\circle*{0.2}} 
\put(   43.30000,  19.96690){\circle*{0.2}} 
\put(   42.59289,  19.67401){\circle*{0.2}} 
\put(   41.88579,  19.96690){\circle*{0.2}} 
\put(   43.59289,    .77452){\circle*{0.2}} 
\put(   45.00711,    .77452){\circle*{0.2}} 
\put(   29.47619,  10.00000){\circle*{0.2}} 
\put(   27.47619,  10.00000){\circle*{0.2}} 
\put(   29.18329,  10.70711){\circle*{0.2}} 
\put(   27.76908,   9.29289){\circle*{0.2}} 
\put(   28.47619,  11.00000){\circle*{0.2}} 
\put(   28.47619,   9.00000){\circle*{0.2}} 
\put(   27.76908,  10.70711){\circle*{0.2}} 
\put(   29.18329,   9.29289){\circle*{0.2}} 
\end{picture}		   
\caption{{\it Left:} The decompositions $\mathbb{E}_k={\bf E}\oplus {\bf E}^\perp=
{\bf E}\oplus {\bf E}'\oplus {\bf E}''={\mathcal E}\oplus {\bf E}''$.
{\it Centre:} A one-shell $D_8$-cluster $\mathcal{C}$. 
{\it Right:} A fragment of the set $\mathcal{Q}$ defined by $\mathcal{C}$.} 
\end{figure}
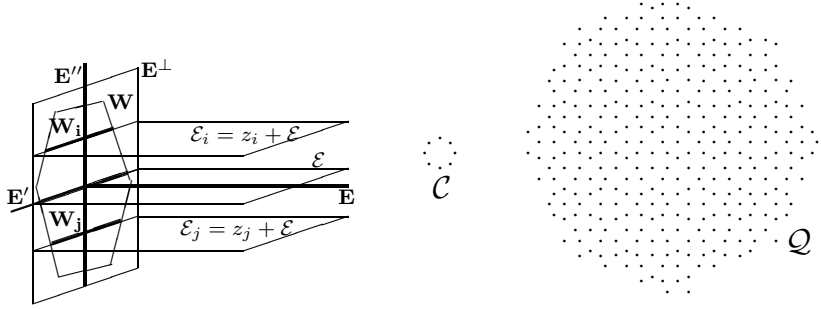

In view of this result, there is a $G$-invariant subspace ${\bf E}'$ and 
a subspace ${\bf V}$ such that ${\bf E}^\perp ={\bf E}'\oplus {\bf V}$, 
$\pi ^\perp (\mathbb{L})\cap {\bf E}'$ is a $\mathbb{Z}$-module dense in ${\bf E}'$, and 
$\pi ^\perp (\mathbb{L})\cap {\bf V}$ is a lattice in ${\bf V}$.
Since ${\bf E}^\perp $ and ${\bf E}'$ are $G$-invariant, the orthogonal 
complement (see figure 1)
\begin{equation}
{\bf E}''=\{ x\in {\bf E}^\perp \ |\ \langle x,y\rangle 
=0\ {\rm for\ all\ } y\in {\bf E}' \} 
\end{equation} 
of ${\bf E}'$ in ${\bf E}^\perp $ is also a $G$-invariant space.
For each $x\in \mathbb{E}_k$ there exist $\pi x \in {\bf E}$, 
$x'\in {\bf E}'$ and $x''\in {\bf E}''$ uniquely determined such that 
$x=\pi x +x'+x''$. The mappings
\begin{equation}\begin{array}{l}
\pi ':\mathbb{E}_k\longrightarrow \mathbb{E}_k:\ x\mapsto \pi 'x=x'\\[1mm]
\pi '':\mathbb{E}_k\longrightarrow \mathbb{E}_k:\ x\mapsto \pi ''x=x''
\end{array}
\end{equation}
are the orthogonal projectors corresponding to ${\bf E}'$ and ${\bf E}''$.\\[3mm]
{\bf Theorem 7.} {\it The $\mathbb{Z}$-module  $\pi (\mathbb{L})$ is either discrete or 
dense in ${\bf E}$.}\\

{\bf Proof.} In view of theorem 6, there is a $G$-invariant subspace $\mathcal{V}_1$
and a subspace $\mathcal{V}_2$ such that ${\bf E}=\mathcal{V}_1\oplus \mathcal{V}_2$,
$\pi (\mathbb{L})\cap \mathcal{V}_1$ is dense in $\mathcal{V}_1$, and 
$\pi (\mathbb{L})\cap \mathcal{V}_2$ is a lattice in $\mathcal{V}_2$.
Since the representation of $G$ in ${\bf E}$ is irreducible we must have either
$\mathcal{V}_1=\{ 0\}$ or $\mathcal{V}_1={\bf E}$.\qquad  \rule{2mm}{2mm}\\

Let $\tilde{\pi }=\pi +\pi '$,  
$\mathcal{E}=\tilde{\pi }(\mathbb{E}_k)={\bf E}\oplus {\bf E}'$,
and let $p:\mathcal{E}\longrightarrow {\bf E}$, \ 
$p':\mathcal{E}\longrightarrow {\bf E}'$ be the restrictions of $\pi $ and $\pi '$
to $\mathcal{E}$.\\[3mm]
{\bf Theorem 8.} {\it The $\mathbb{Z}$-module $\mathcal{L}=\tilde{\pi }(\mathbb{L})$ 
            is a lattice in $\mathcal{E}$.}\\

{\bf Proof.}
If $\pi ^\perp (\mathbb{L})$ is discrete in ${\bf E}^\perp $ then ${\bf E}'=\{ 0\}$,
and in view of theorem 4 the projection $\mathcal{L}=\pi (\mathbb{L})$ of 
$\mathbb{L}$ on $\mathcal{E}={\bf E}$ is a lattice in $\mathcal{E}$.

If $\pi ^\perp (\mathbb{L})$ is not discrete in ${\bf E}^\perp $ then
for any $\rho \in (0,\infty )$ the dimension of the subspace $V_\rho $
generated by the set 
$\{ \pi ^\perp x\ |\ x\in \mathbb{L},\ ||\pi ^\perp x||<\rho \}$ is greater 
than or equal to one. More than that,  $\varrho \leq \rho '\Longrightarrow
V_\rho \subset V_{\rho '}$, and there is $\rho _0\in (0,\infty )$
such that $V_\rho =V_{\varrho _0}$ for any $\rho \leq \rho _0$.
We have ${\bf E}'=V_{\rho _0}$. Since
\[ \left. \begin{array}{l}
x\in \mathbb{L}\\
\pi ^\perp x\not\in  {\bf E}'
\end{array}\right\}\quad \Longrightarrow \quad ||\pi ''x||>\rho _0 \]
$\pi ''(\mathbb{L})$ is a lattice in ${\bf E}''$. 
From theorem 4 it follows that $\mathcal{L}$
is a lattice in $\mathcal{E}$.
\qquad  \rule{2mm}{2mm}\\[3mm]
{\bf Theorem 9.} {\it If $\pi (\mathbb{L})$ is dense in ${\bf E}$ then 
\begin{equation}
\begin{array}{ccccc}
{\bf E}& \stackrel{p}\longleftarrow & \mathcal{E}
& \stackrel{p'}\longrightarrow & {\bf E}' \\
&&\cup && \\
&& \mathcal{L} &&
\end{array}
\end{equation} 
is a cut and project scheme.}\\

{\bf Proof.} From the relation 
$p'(\mathcal{L})=\pi '((\pi +\pi ')(\mathbb{L}))=\pi '(\mathbb{L})$ and the definition
of ${\bf E}'$ it follows that $p'(\mathcal{L})$ is dense in ${\bf E}'$.

Let $\tilde x,\tilde y\in \mathcal{L}$ with $p\tilde x=p \tilde y$, and let 
$x,y\in \mathbb{L}$ be such that $\tilde x=\tilde \pi x$ and $\tilde y=\tilde \pi y$.
From $p\tilde x=p \tilde y$ it follows $\pi x=\pi y$, whence $x-y\in {\bf E}^\perp $.
Since $\mathbb{L}\cap {\bf E}''$ contains a basis of ${\bf E}''$ and  
$\mathbb{L}\cap {\bf E}'=\{ 0\}$ we must have $x-y\in {\bf E}''$, whence
$\tilde x=\tilde y$. Therefore $p$ restricted to $\mathcal{L}$ is one-to-one.
\qquad  \rule{2mm}{2mm}\\

Let ${\bf W}=\pi ^\perp (\mathbb{W})$ be the projection of the hypercube
\begin{equation}
\mathbb{W}=v+\{ \ (x_i)_{1\leq i\leq k}\ |\ 0\leq x_i\leq \kappa \ \}
\end{equation}
where the translation $v\in {\bf E}'$ 
is such that no point of $\pi ^\perp (\mathbb{L})$
belongs to $\partial {\bf W}$.\\[3mm] 
{\bf Theorem 10.} {\it If $\pi (\mathbb{L})$ is dense in ${\bf E}$ then the set
\begin{equation}
{\mathcal Q}=\left\{ \left. \pi  x\ \right| \ 
x\in \mathbb{L},\ \pi ^\perp x\in {\bf W} \right\}
\end{equation}
is a multi-component model set.}\\

{\bf Proof.}
Let $\Theta =\{ \theta _i\ |\ i\in \mathbb{Z}\}$ be a subset of $\mathbb{L}$
such that $\pi ''(\mathbb{L})=\pi ''(\Theta )$ and 
$\pi ''\theta _i\not=\pi ''\theta _j$ for $i\not=j$. The lattice $\mathbb{L}$
is contained in the union $\bigcup_{i\in \mathbb{Z}}\mathcal{E}_i$ of all the
cosets $\mathcal{E}_i=\theta _i+\mathcal{E}$ of $\mathcal{E}$ in $\mathbb{E}_k$.
The lattice ${\bf L}=\mathbb{L}\cap \mathcal{E}$ is a sublattice of 
$\mathcal{L}$. Since $\mathbb{L}\cap \mathcal{E}_i=\theta _i+{\bf L}$ the set
$\mathcal{L}_i=\tilde{\pi }(\mathbb{L}\cap \mathcal{E}_i)
=\tilde{\pi }\theta _i+{\bf L}$
is a coset of ${\bf L}$ in $\mathcal{L}$ for any $i\in \mathbb{Z}$.
Since $\pi ''(\mathbb{L})$ is discrete in ${\bf E}''$, the intersection (see figure 1)
\[ {\bf W}_i={\bf W}\cap {\mathcal E}_i={\bf W}\cap \pi ^\perp (\mathcal{E}_i)
=\pi ^\perp (\mathbb{W}\cap {\mathcal E}_i)\subset \pi ''\theta _i+{\bf E}'\]
is non-empty only for a finite number of cosets $\mathcal{E}_i$.
By changing the indexation of the elements of $\Theta $
if necessary, we can assume that the subset
${\mathcal W}_i=\pi '({\bf W}_i)=\pi '(\mathbb{W}\cap {\mathcal E}_i)$ of ${\bf E}'$
has a non-empty interior only for $i\in \{ 1,...,m\}.$ 
The `polyhedral' sets ${\mathcal W}_i$ satisfy the conditions (i)-(iii) from
definition 3, and
\begin{equation}\label{defQ}
{\mathcal Q} =\bigcup_{i=1}^m
\left\{ \pi x\ \left| \ x\in {\mathcal L}_i , \ \pi ' x\in {\mathcal W}_i
\right.\right\}.
\end{equation} 
\rule{2mm}{2mm}\\

The set $\mathcal{Q}$ is a union of interpenetrating copies of the starting 
cluster ${\mathcal C}$, most of them only partially occupied. 
For each point $\pi x\in {\mathcal Q}$ the set of all the arithmetic neighbours 
of $\pi x$
\[ \left\{ \pi y \left|  
\begin{array}{l}
y\in \{ x+\kappa \varepsilon _1,...,x+\kappa \varepsilon _k,
x-\kappa \varepsilon _1,...,x-\kappa \varepsilon _k\}\\  
\pi ^\perp y\in {\bf W}
\end{array}\right. \right\}\]
is contained in the translated copy 
\[ \{ \pi x +e_1,...,\pi x+e_k,\pi x -e_1,...,\pi x-e_k\}=\pi x +{\mathcal C} \]
of the $G$-cluster ${\mathcal C}$.
An algorithm and some software for generating such patterns is available
via internet \cite{C}. Some examples are presented in figures 1 and 2.
In each case, the quasiperiodic pattern is a packing of partially occupied
copies of the corresponding cluster. One can remark that the occupation of
these copies seem to be very low in the case of a multi-shell cluster.

\begin{figure}
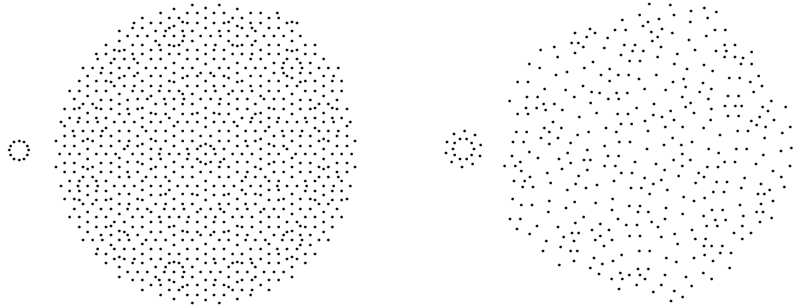

\setlength{\unitlength}{1.3mm}
%
		   
\caption{{\it Left:} A one-shell $D_{12}$-cluster and a fragment of the corresponding 
quasiperiodic set. 
{\it Right:} A fragment of the quasiperiodic set defined by a two-shell $D_{10}$-cluster.} 
\end{figure}

The number $\alpha $ is called a {\it scaling factor} of $\mathcal{Q}$ if there is
$y\in {\bf E}$ such that $\mathcal{Q}$ is invariant under the affine similarity \cite{MPP}
\begin{equation} 
A:{\bf E}\longrightarrow {\bf E}\qquad Ax=y+\alpha (x-y)
\end{equation} 
that is, if $A(\mathcal{Q})\subset \mathcal{Q}$.
In this case we say \cite{MPP} that $y$ is an {\it inflation center} corresponding 
to $\alpha $,
and $A$ is a {\it self-similarity} of $\mathcal{Q}$. The definition (\ref{defQ}) 
offers some
facilities \cite{C,K} in the study of the self-similarities of $\mathcal{Q}$.

In view of the theorem 3, each quasiperiodic set defined by the above algorithm 
is a Delone set, has a well-defined density and is pure point diffractive.

\section{An example}

In order to illustrate the algorithm presented in the previuos section we consider
the dihedral group 
$D_{10}=\left< a,\; b\ \left|\ a^{10}=b^2=(ab)^2=e\right. \right>$, 
the two-dimensional representation
\[ \begin{array}{l}
a(\alpha ,\beta )\!=\!\left(\alpha \cos \frac{\pi }{5}-\beta \sin \frac{\pi }{5},
\alpha \sin \frac{\pi }{5}+\beta \cos \frac{\pi }{5}\right)\\[2mm]
b(\alpha ,\beta )\!=\!(\alpha ,-\beta ) 
\end{array} \]  
and the $D_{10}$-cluster ${\mathcal C}=D_{10}(1,0)$ generated by the set 
$S=\{ (1,0) \}.$ 

The action of $a$ and $b$ on 
${\mathcal C}$ 
generate the orthogonal representation of $D_{10}$ in $\mathbb{E}_5$ 
\begin{equation} \begin{array}{l} 
a(x_1,x_2,x_3,x_4,x_5)=(-x_3,-x_4,-x_5,-x_1,-x_2)\\ 
b(x_1,x_2,x_3,x_4,x_5)=(x_1,x_5,x_4,x_3,x_2). 
\end{array}  
\end{equation} 
The matrices of the projectors corresponding to ${\bf E}$, ${\bf E}'$, 
${\bf E}''$ are
$\pi \!=\!{\mathcal M}\left( \frac{2}{5},-\frac{\tau '}{5},-\frac{\tau }{5}\right)$,
$\pi '\!=\!{\mathcal M}\left( \frac{2}{5},-\frac{\tau }{5},-\frac{\tau '}{5}\right)$,
$\pi ''\!=\!\mathcal {M}\left( \frac{1}{5},\frac{1}{5},\frac{1}{5}\right)$,
where  $\tau =(1+\sqrt{5})/2$, \ $\tau ' =(1-\sqrt{5})/2$ and
\begin{equation} 
{\mathcal M}(\alpha ,\beta ,\gamma )=\left( \begin{array}{rrrrr} 
\alpha &\beta &\gamma &\gamma &\beta  \\ 
\beta & \alpha &\beta &\gamma &\gamma \\  
\gamma &\beta & \alpha &\beta &\gamma \\ 
\gamma &\gamma &\beta &\alpha &\beta  \\ 
\beta &\gamma &\gamma &\beta &\alpha  
	     \end{array} \right) . 
\end{equation} 
In this case $\kappa =\sqrt{5/2}$, and the projection of 
$\mathbb{L}=\kappa \mathbb{Z}^5$ on the space
\[ 
\mathcal{E}={\bf E}\oplus{\bf E}'
=\{ (x_1,x_2,...,x_5)\ |\ x_1+x_2+...+x_5=0\}
\]
is the lattice 
$\mathcal{L}=\mathbb{Z}\chi _1+\mathbb{Z}\chi _2+\mathbb{Z}\chi _3+\mathbb{Z}\chi _4$,
where $\chi _j=(\pi +\pi ')(\kappa \varepsilon _j).$
If ${\bf W}=\pi ^\perp (\mathbb{W})$ is the projection on 
${\bf E}^\perp ={\bf E}'\oplus {\bf E}''$ of a hypercube
\[ \mathbb{W}=v+\{ (x_1,x_2,x_3,x_4,x_5)\ |\ 0\leq x_j\leq \kappa \ {\rm for\ 
any\ } j\} \]
with $v\in {\bf E}'$ chosen such that no point of $\pi ^\perp (\mathbb{L})$ belongs to 
$\partial {\bf W}$ then
\begin{equation}
\mathcal{Q}=\left\{ \left. \pi x\ \right| \ 
x\in \mathbb{L},\ \pi ^\perp x\in {\bf W}\right\} 
\end{equation}
is the set of all the vertices of a {\em rhombic Penrose tiling} \cite{K}.

The lattice $\mathbb{L}$ is contained in the union 
$\bigcup_{j\in \mathbb{Z}}\mathcal{E}_j$ of subspaces
\[  {\mathcal E}_j=\{ (x_1,x_2,...,x_5)\ 
|\ x_1+x_2+...+x_5=j\kappa \}=\theta _j+\mathcal{E}\]
where $\theta _j=(j\kappa ,0,0,0,0)\in \mathbb{L}$. 
Since $\mathbb{L}\cap \mathcal{E}_j=\theta _j+{\bf L}$, the set 
\begin{equation}
{\mathcal L}_j=\tilde \pi (\mathbb{L}\cap {\mathcal E}_j)=
\tilde \pi \theta _j+{\bf L}=j\chi _1+{\bf L}
\end{equation}
is a coset of ${\bf L}=\mathbb{L}\cap \mathcal{E}$ 
in $\mathcal{L}$, for any $j\in \mathbb{Z}$.

The set $\mathbb{W}\cap {\mathcal E}_j$ is non-empty only for $j\in \{ 0,1,2,3,4,5\}$,
but ${\mathcal W}_j=\pi '(\mathbb{W}\cap {\mathcal E}_j)$ has non-empty interior 
only for $j\in \{ 1,2,3,4\}$. 
Let $\mathcal{P} \subset {\bf E}'$ be the set of all the points lying 
inside or on the boundary of the regular pentagon with the vertices
$\pi '(\kappa ,0,0,0,0)$, $\pi '(0,\kappa ,0,0,0)$, ..., $\pi '(0,0,0,0,\kappa ).$
One can remark that $\mathcal{W}_1=v+\mathcal{P} $, 
$\mathcal{W}_2=v-\tau \mathcal{P}$, 
$\mathcal{W}_3=v+\tau \mathcal{P}$,
$\mathcal{W}_4\!=\!v\!-\!\mathcal{P}$,  
and express the pattern $\mathcal{Q}$ as a multi-component model set (see figure 3)
\begin{equation}
\mathcal{Q}=\bigcup_{i=1}^4
\left\{ p\, x\ \left| \ x\in {\mathcal L}_i , \ p'\, x\in \mathcal{W}_i
\right.\right\} .
\end{equation}
This definition is directly related to de Bruijn's definition \cite{M}.

\begin{figure}
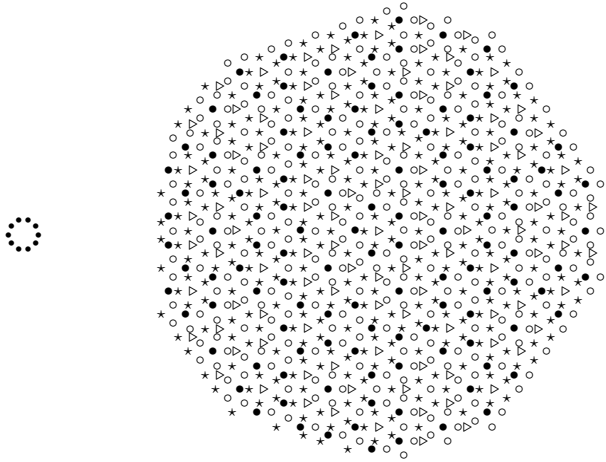
 
\setlength{\unitlength}{2mm}  
  
\caption{The cluster $\mathcal{C}=D_{10}(1,0)$ and a fragment of the 
quasiperiodic pattern defined by this cluster. The positions of the points 
are indicated by using the symbols {\tiny $\bullet $}, {\tiny $\circ $},
{\tiny $\star $} and {\tiny $\triangleright $} in order to distinguish 
the four components of this model set.} 
\end{figure}	 

The particular case of the icosahedral group, very important for quasicrystal
physics, has been presented (with direct proofs) in \cite{C3}.

\section{Concluding remarks}

Quasicrystal structure analysis comprises the determination on an atomic scale
of the short-range order (atomic arrangement inside the structural building unit)
as well as the long-range order (the way the structural building units are arranged 
on the long scale). The rational approximants (periodic crystals with the same
building unit as the considered quasicristal) provide a powerful way to determine
the short-range order, but the description of the long-range order is still a
major problem \cite{St}. Only in the case of a few decagonal and icosahedral phases,
the electron microscopic and diffraction data have allowed us to have an idea
about the structure of the building unit and long-range ordering. 

The atomic structure can not be extracted directly from the experimental data. 
One has to postulate a structure and to compare the forecasts
with the electron microscopic and diffraction  data. 
There exist several attempts in this direction:
\begin{itemize}
\item[-] 
Elser \& Henley \cite{E1} and  Audier \& Guyot \cite{A} 
have obtained  models for icosahedral quasicrystals 
by decorating the Ammann rhombohedra occuring in a tiling of the 3D space
defined by projection \cite{E,K}. 
\item[-] 
In his quasi-unit cell picture Steinhardt 
\cite{SJ} has shown (following an idea of Petra Gummelt \cite{Gum}) that the atomic 
structure can be described entirely by using a single repeating cluster which overlaps
(shares atoms with) neighbour clusters. The model is determined by the overlap 
rules and the atom decoration of the unit cell. 
\item[-]
Some important models have been
obtained by Yamamoto \& Hiraga \cite{Y1,Y2}, 
Katz \& Gratias \cite{K1}, Gratias, Puyraimond and Quiquandon \cite{GPQ} 
by using the section method in a
6D superspace decorated with several polyhedra (acceptance domains).
\item[-]
Janot and de Boissieu \cite{J} have shown that a model of icosahedral quasicrystal
can be generated recursively by starting from a pseudo-Mackay cluster 
and using some inflation rules. 
\end{itemize}

Most information about the type of quasiperiodic long-range order is in the very
weak reflections. The number of Bragg reflections we can observe is too small 
for an accurate structure description \cite{St}.
The experimental devices allow us to obtain diffraction patterns and to 
have a direct view on some small fragments of the quasicrystal. These data provide 
easy access to the symmetry group $G$ and allows us to look for an adequate  
cluster $\mathcal{C}$. The pattern $\mathcal {Q}$ obtained by using our algorithm is
exactly defined and has the remarkable mathematical properties of the patterns obtained 
by projection. Each point of our pattern (without exception) is
the center of a more or less occupied copy of $\mathcal {C}$, but 
unfortunately, in the case of complex clusters 
the occupation is extremely low for most of the points of $\mathcal{Q}$ . 
Therefore, our discrete quasiperiodic sets can not be used directly
in the description of atomic positions in quasicrystals.

The cluster $\mathcal {C}$ can be regarded as a covering cluster, and $\mathcal{Q}$ 
as a quasiperiodic set which can be covered by partially occupied copies of a single
cluster. This kind of covering is different from the covering of Penrose tiling by
a decorated decagon \cite{Gum} proposed by Gummelt in 1996 or the coverings of discrete 
quasiperiodic sets presented by Kramer, Gummelt, G\" ahler {\it et. al.} 
in \cite{KP}. In our case the generating cluster $\mathcal{C}$ is a finite set of 
points and the quasiperiodic pattern is obtained by projection. In \cite{Gum,KP} the covering
clusters  are congruent overlapping polytopes (with an asymetric decoration) and the
structure is generated by imposing  certain overlap rules which restrict the possible 
relative positions and orientations of neighbouring clusters. 
When the theory from \cite{Gum,KP} is applied to quasicrystals, 
atomic positions are assigned to the covering clusters.

There are some indications that stable clusters are smaller than the basic 
structural units seen on electron microscopic images, and it is believed 
that larger clusters automatically introduce disorder. The defects occuring in the tilings
constructed from electron microscopic images show that a certain amount of 
disorder either in glue atoms or in the clusters seems to be unavoidable \cite{GPQ}.
In the case of Gummelt's approach, the transition from perfect to random 
quasicrystalline order is obtained by passing to relaxed overlap rules \cite{RG}.
The frequency of occurrence of fully occupied clusters in our
quasiperiodic patterns can be increased by a certain relaxation in the use
of strip projection method, but this leads to some defects. In order to
correct these defects one has to elliminate some points from interpenetrating
clusters if they become too close.\\[5mm] 
{\bf Acknowledgement.} This research was supported by the grant CEEX 582/2005.


\begin{thebibliography}{10}

\bibitem{A}
M. Audier and P. Guyot, $Al_4Mn$ quasicrystal atomic structure, diffraction
         data and Penrose tiling, Phil. Mag. B 53 (1986) L43--L51.

\bibitem{B} 
M. Baake and R. V. Moody, Multi-component model sets and invariant densities, 
       in {\it Proc. Int. Conf. Aperiodic' 97 (Alpe d'Huez, 27-31 August, 1997)},
          Eds. M. de Boissieu, J.-L. Verger-Gaugry and R. Currat 
           (World Scientific, Singapore, 1999)   pp.~9--20.

\bibitem{B1}
M. Baake and R. V. Moody, Weighted Dirac combs with pure point diffraction,
        J. reine angew. Math. 573 (2004) 61--94.

\bibitem{B2}
M. Baake, R. V. Moody, C. Richard and B. Sing, Which distributions of matter 
        diffract ? - Some answers, in {\it Quasicrystals: Structure and 
        Physical Properties}, Ed. H.-R. Trebin (Wiley-VCH, Berlin, 2003) pp.~188-207.


\bibitem{C1} 
N. Cotfas and J.-L. Verger-Gaugry, A mathematical construction
         of $n$-dimensional quasicrystals starting from $G$-clusters,
          J. Phys. A: Math. Gen. 30 (1997) 4283-4291.

\bibitem{C2}
N. Cotfas, Permutation representations defined by $G$-clusters with
         application to quasicrystals, 
         Lett. Math. Phys. 47 (1999) 111-123.

\bibitem{C3}
N. Cotfas, Icosahedral multi-component model sets,
         J. Phys. A: Math. Gen. 37 (2004) 3125-3132.

\bibitem{C}
N. Cotfas, {\bf http:}\verb#//fpcm5.fizica.unibuc.ro/~ncotfas# \ .

\bibitem{D}
D. Descombes, {\it El\' ements de Th\' eorie des Nombres}
          (PUF, Paris, 1986) pp 54-59.

\bibitem{E} 
V. Elser, The diffraction pattern of projected structures,
        Acta Cryst. A  42 (1986) 36-43.

\bibitem{E1}
V. Elser and C. L. Henley, Crystal and quasicrystal structures in Al-Mn-Si 
         alloys, Phys. Rev. Lett. 55 (1985) 2883-2886.

\bibitem{GPQ}
D. Gratias, F. Puyraimond and M. Quiquandon, Atomic clusters in icosahedral
        F-type quasicrystals, Phys. Rev. B 63 (2000) 024202.

\bibitem{Gum}
P. Gummelt, Penrose Tilings as coverings of congruent decagons, 
         Geometriae Dedicata 62  (1996) 1-17. 

\bibitem{H1}
A. Hof, On diffraction by aperiodic structures, Commun. Math. Phys.
        169 (1995) 25-43.

\bibitem{H2}
A. Hof, Diffraction by aperiodic structures, in {\it The Mathematics of 
         Long-Range Aperiodic Order}, ed. R. V. Moody 
         (Kluwer Acad Publ., Dordrecht, 1997) pp 239-286.

\bibitem{J}
C. Janot and M. de Boissieu, Quasicrystals as a hierarchy of clusters,
         Phys. Rev. Lett. 72 (1994) 1674-1677.

\bibitem{KKL}
P. A. Kalugin, A. Y. Kitayev and L.S. Levitov,  6-dimensional properties
      of AlMn alloy, J. Physique Lett. 46 (1985) L601-L607.

\bibitem{K}
A. Katz and M. Duneau, Quasiperiodic patterns and icosahedral symmetry,
          J. Phys. (France) 47 (1986) 181-196.

\bibitem{K1}
A. Katz and D. Gratias, A geometric approach to chemical ordering in
      icosahedral structures, J. Non-Cryst. Solids 153\&154 
      (1993) 187-195.

\bibitem{KN}
P. Kramer and R. Neri, On periodic and non-periodic space fillings of 
         $\mathbb{E}^m$ obtained by projection, Acta Crystallogr. A 40 
         (1984) 580-587.

\bibitem{KP}
P. Kramer and Z. Papadopolos (eds.), {\it Coverings of Discrete Quasiperiodic Sets.
Theory and Applications to Quasicrystals} (Springer-Verlag, Berlin, 2003).

\bibitem{L1}
J. C. Lagarias, Meyer's concept of quasicrystal and quasiregular sets,
     Commun. Math. Phys. 179 (1996) 365-376.

\bibitem{L2}
J. C. Lagarias, Mathematical quasicrystals and the problem of diffraction, in
      {\it Directions in Mathematical Quasicrystals}, eds. M. Baake and R. V. Moody, 
      CRM Monograph Series vol. 13 (AMS, Providence, RI, 2000) pp 61-93.

\bibitem{LMS1}
J.-Y. Lee, R.~V. Moody, and B. Solomyak, Pure point dynamical and diffraction
spectra, Annales. H. Poincare 3 (2002), 1003–1018.

\bibitem{LMS2}
J.-Y. Lee, R.~V. Moody, and B. Solomyak, 
Consequences of Pure Point Diffraction Spectra for Multiset Substitution Systems, 
Discrete Comput. Geom. 29 (2003) 525-560.

\bibitem{MPP}
Z. Mas\' akov\' a, J. Patera J and E. Pelantov\' a, Inflation centers of the cut and 
        project quasicrystals, J. Phys. A: Math. Gen. 31 (1998) 1443-1453.

\bibitem{Me}
Y. Meyer, {\it Algebraic Numbers and Harmonic Analysis} (North-Holland,
         Amsterdam, 1972) pp. 48-49.

\bibitem{M} 
R. V. Moody, Meyer sets and their duals, in {\it The Mathematics of Long-Range 
       Aperiodic Order}, ed. R. V. Moody (Kluwer, Dordrecht, 1997) pp 411-412.

\bibitem{RG}
M. Reichert and F. G\" ahler, Cluster model of decagonal tilings, 
       Phys. Rev. B 68 (2003) 214202.

\bibitem{Sch1}
M. Schlottmann, Cut-and-project sets in locally compact Abelian groups, in
        {\it Quasicrystals and Discrete Geometry}, ed. J. Patera, pp. 247--264,
        Fields Institute Monographs vol. 10 (AMS, Providence, RI, 1998) pp 247-264.

\bibitem{Sch2}
M. Schlottmann, Generalized model sets and dynamical systems, in 
      {\it Directions in Mathematical Quasicrystals}, eds. M. Baake
         and R. V. Moody, CRM Monograph Series vol. 13 (AMS, 
         Providence, RI, 2000) pp 143-159.

\bibitem{S} 
M. Senechal, {\it Quasicrystals and Geometry}
         (Cambridge University Press, Cambridge, 1995) pp 264-266.

\bibitem{St}
W. Steurer, Quasicrystal structure analysis, a never-ending story ?,
       J. Non-Cryst. Solids 334\&335 (2004) 137-142.

\bibitem{SJ}
P. J. Steinhardt and H.-C. Jeong, A simpler approach to Penrose tiling 
       with implications for quasicrystal formation, Nature 382 (1996) 433-435.

\bibitem{Y1} 
A. Yamamoto and K. Hiraga, Structure of an icosahedral Al-Mn quasicrystal,
       Phys. Rev. B 37 (1988) 6207-6214.

\bibitem{Y2} 
A. Yamamoto, Ideal structure of icosahedral Al-Cu-Li quasicrystals,
        Phys. Rev. B 45 (1992) 5217-5227.

\end{thebibliography}
\end{document}